\documentclass[letter]{aa}  

\usepackage{natbib}
\bibpunct{(}{)}{;}{a}{}{,} % to follow the A&A style
\usepackage{graphics}   % standard graphics specifications
\usepackage{graphicx}   % alternative graphics specificati.jpgons
\usepackage{epstopdf}
\usepackage{longtable}   % helps with long table options
\usepackage{url}      % for on-line citations
\usepackage{bm}        % special 'bold-math' package
\usepackage[varg]{txfonts}
\usepackage{pdflscape}
\usepackage{supertabular}
\usepackage{hyperref}   %link

\hypersetup{
    bookmarks=true,         % show bookmarks bar?
    unicode=false,          % non-Latin characters in Acrobat?s bookmarks
    colorlinks=true,       % false: boxed links; true: colored links
    linkcolor=blue,          % color of internal links (change box color with linkbordercolor)
    citecolor=blue,        % color of links to bibliography
    filecolor=blue,      % color of file links
    urlcolor=blue           % color of external links
}
\usepackage{color}

\begin{document}

\title{Star-planet interactions} 
\subtitle{IV. Possibility of detecting the orbit-shrinking of a planet around a red giant}

\author{Georges Meynet\inst{1}, Patrick Eggenberger\inst{1}, Giovanni Privitera\inst{1}, Cyril Georgy\inst{1}, Sylvia Ekstr{\"o}m\inst{1}, Yann Alibert\inst{2}, and
Christophe Lovis\inst{1}
%\author{Georges Meynet\inst{1}, Patrick Eggenberger\inst{1}, Giovanni Privitera\inst{1}, Cyril Georgy\inst{1}, Sylvia Ekstr{\"o}m\inst{1}, and Yann Alibert\inst{2} 
}

 \authorrunning{Meynet et al.}

 \institute{Geneva Observatory, University of Geneva, Maillettes 51, CH-1290 Sauverny, Switzerland
\and Physikalisches Institut \& Center for Space and Habitability, Universitaet Bern, 3012, Bern, Switzerland
}

\date{Received /
Accepted}
\abstract {The surface rotations of some red giants are so fast that they must have been spun up by tidal interaction with a close companion, either another star, a brown dwarf, or a planet.
We focus here on the case of red giants that are spun up by tidal interaction with a planet. When the distance between the planet and the star decreases, the spin period of the star
%$P_{\rm rot,*}$, 
decreases, the orbital
period of the planet
%, $P_{\rm orb}$, 
decreases, and the reflex motion of the star
%, $\upsilon_{\rm reflex,*}$, 
increases.  
We study the change rate of these three quantities when the circular orbit of a planet of 15 M$_{J}$ that initially orbits a 2 M$_\odot$ star at 1 au shrinks under the action of tidal forces
during the red giant phase.
We use stellar evolution models coupled with computations of the orbital evolution of the planet, which allows us to follow the exchanges of angular momentum between the star and the orbit
in a consistent way. We obtain that
the reflex motion of the red giant star increases by more than 1 m s$^{-1}$ per year in the last $\sim$40 years before the planet engulfment. During this phase, the reflex motion of the star is between
660 and 710 m s$^{-1}$. The spin period of the star increases by more than about 10 minutes per year in the last 3000 y before engulfment. During this period, the spin period of the star is shorter than 0.7 year. During this same period,
the variation in orbital period, which is shorter than 0.18 year, is on the same order of magnitude. Changes in reflex-motion and spin velocities are very small  and thus most likely out of reach of being observed.
The most promising way of detecting this effect is through observations of transiting planets, that is, through{\it } changes of the beginning or end of the transit. 
For the relatively long orbital periods expected around red giants, long observing runs of typically a few years are needed. Interesting star-planet systems that currently are in this stage of orbit-shrinking would be red giants with
fast rotation (above typically 4-5 km s$^{-1}$), a low surface gravity (log $g$ lower than 2), and having a planet at a distance typically smaller than about 0.4 - 1 au, depending on log $g$.  A space mission like PLATO might be of great interest for detecting planets that are on the verge of being engulfed by red giants. The discovery of a few systems, even only one, would provide very interesting clues about the physics of tidal interaction between a red giant and a planet.
%However, even with the appropriate observational facilities, it will be difficult to detect these variations in view of the short period during which they occur. Typically 3000 y correspond to about 10$^{-5}$ the red giant branch lifetime of a 2 M$_\odot$. Thus, for the planets, the situation does not seem very favorable for the detection of such variations in red giant stars. Situations does appear
%more favorable for brown dwarfs.
}


\maketitle

\titlerunning{Stellar rotation and planet orbital evolution}
\authorrunning{Meynet et al.}

\section{Introduction}

\begin{figure*}
\centering
\includegraphics[width=.38\textwidth, angle=0]{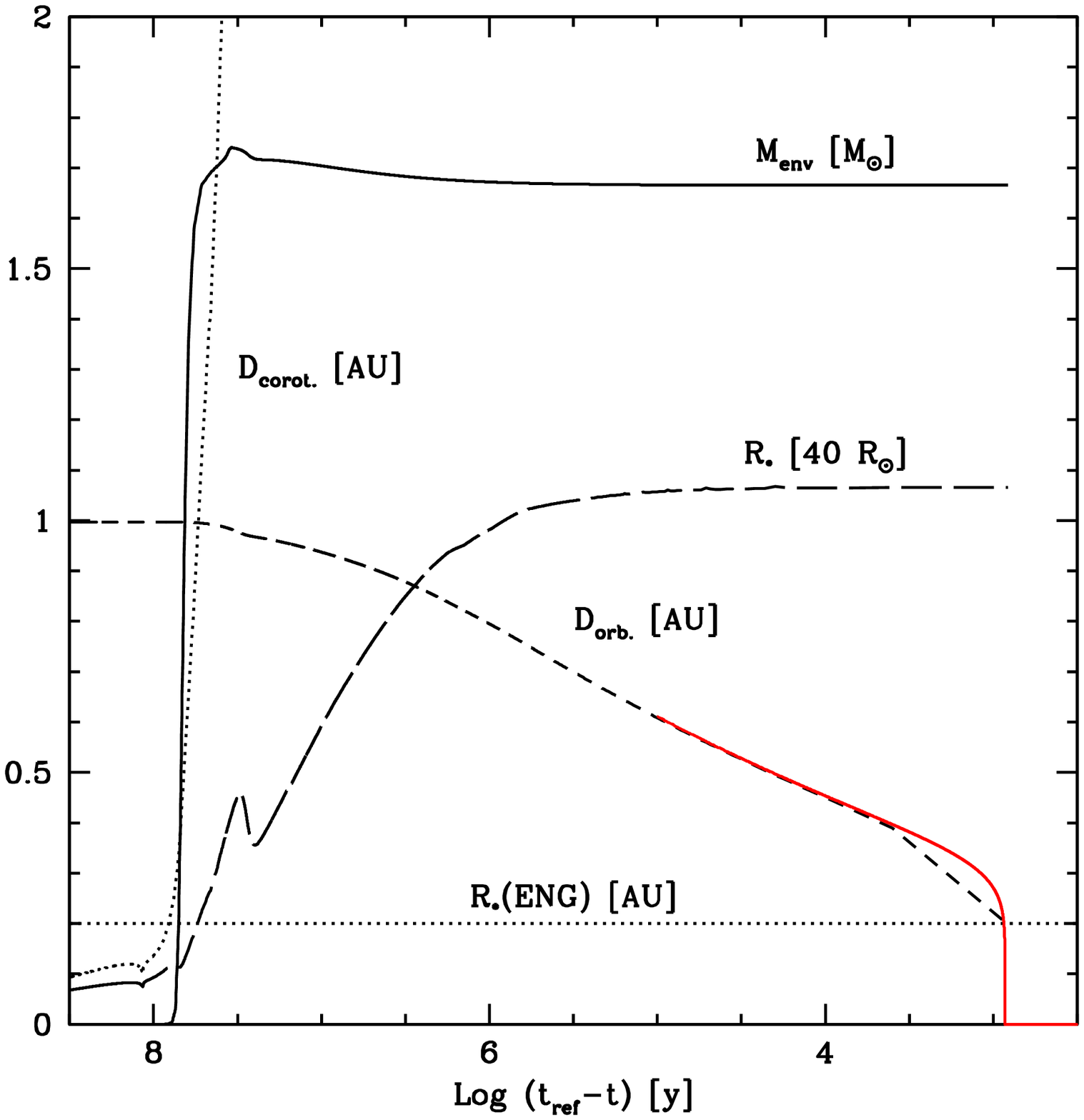}\includegraphics[width=.38\textwidth, angle=0]{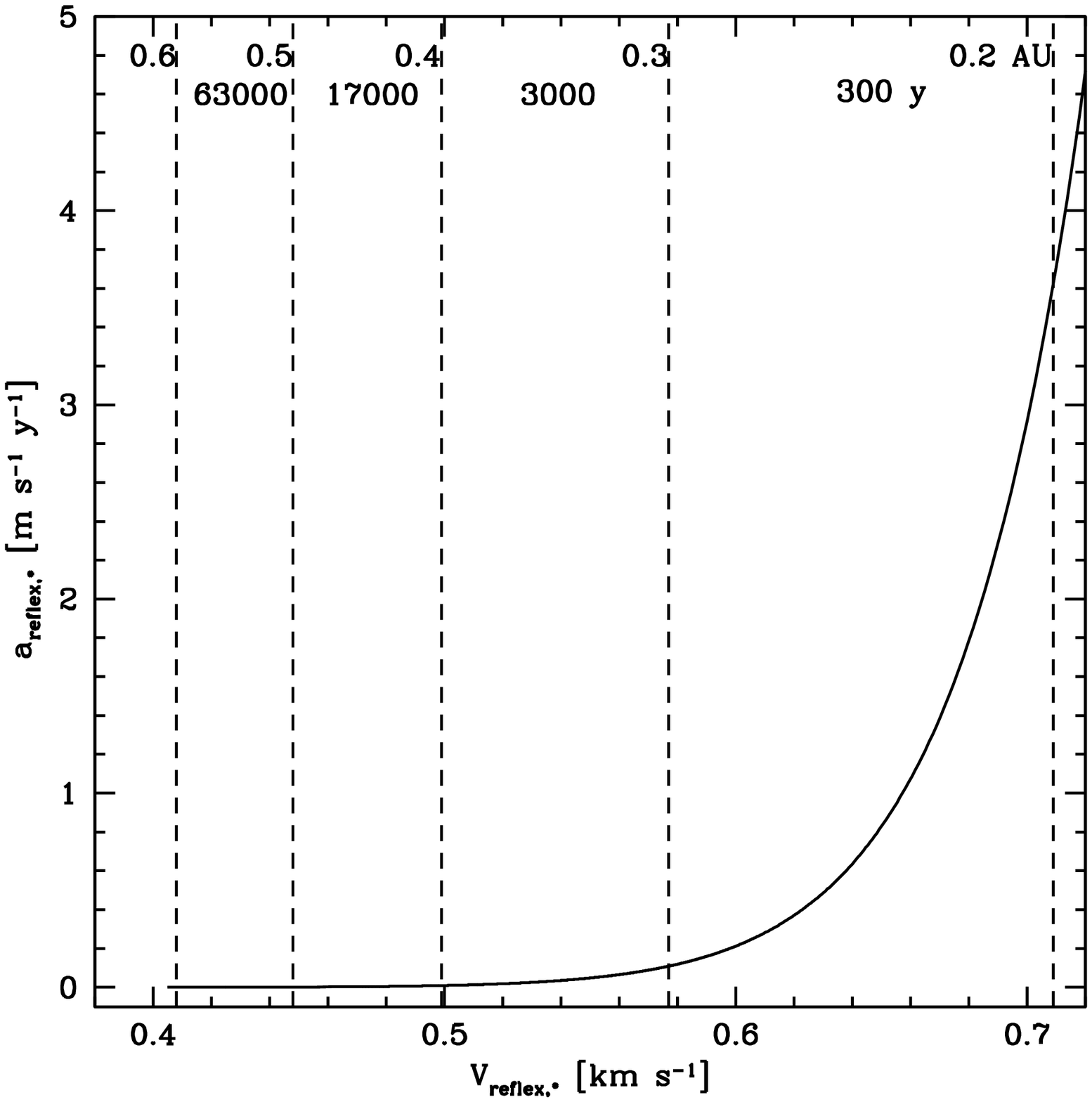}
\includegraphics[width=.38\textwidth, angle=0]{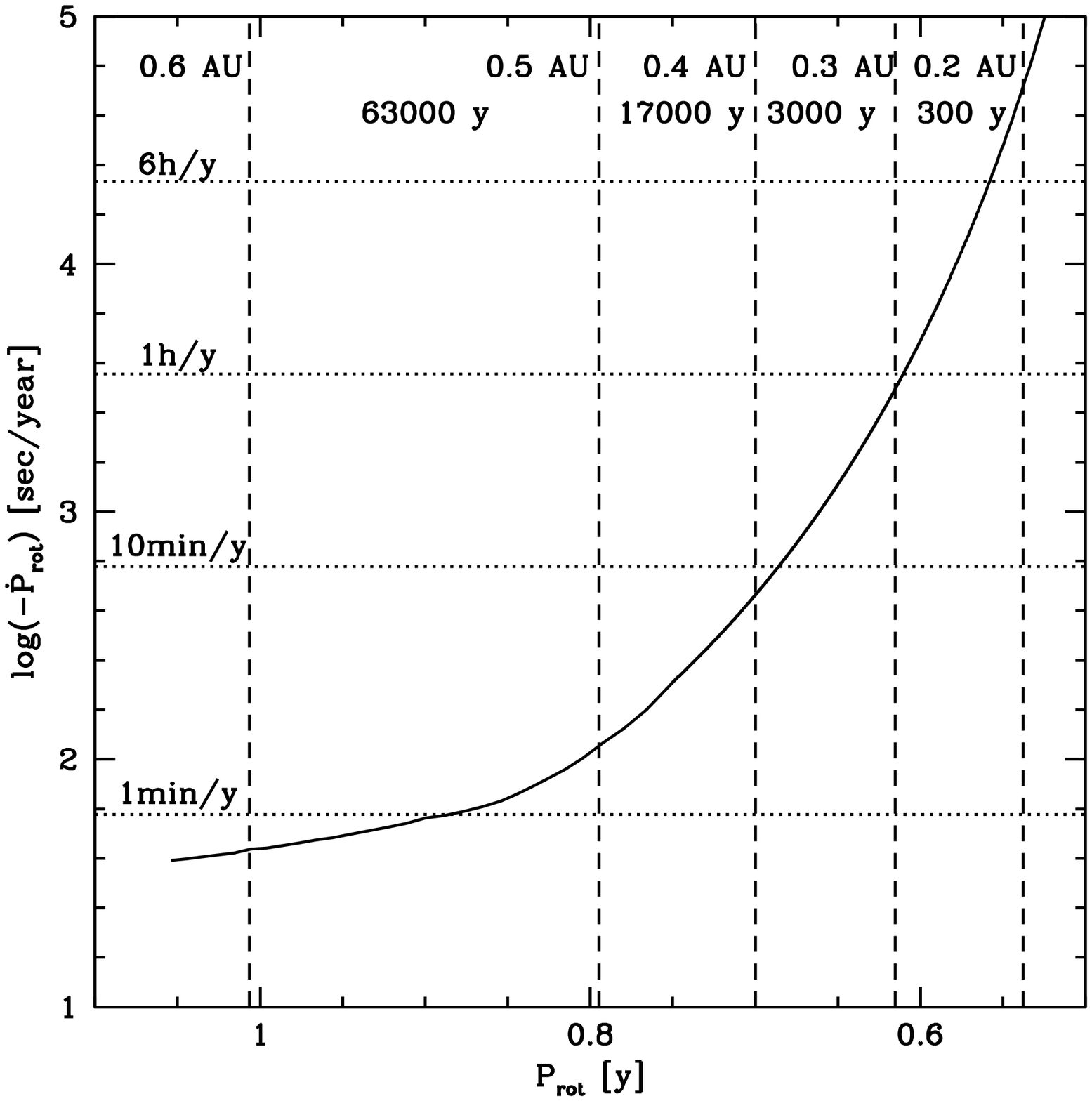}\includegraphics[width=.38\textwidth, angle=0]{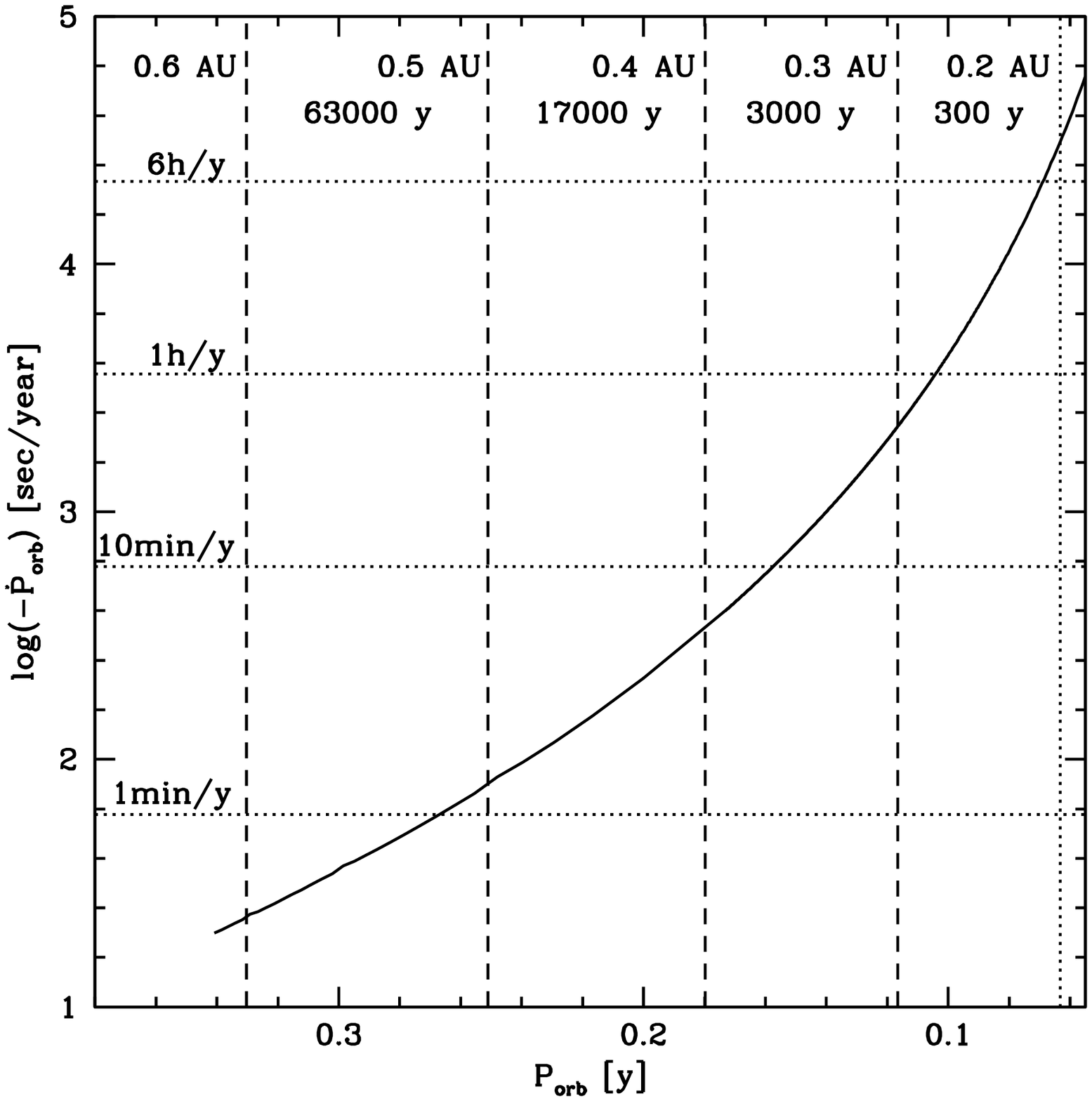}
\caption{{\it Upper left panel:}
Evolution as a function of time of a few relevant quantities during the last 100 Myr before engulfment for a 15 M$_J$ planet orbiting a 2 M$_\odot$ star ($\Omega_{ini}/\Omega_{crit}$=0.1) at an initial distance equal to 1 au. The horizontal axis is the logarithm of the difference between a reference time ($t_{ref}$) taken here equal to 1.29018e+09 years and the age of the star. 
%We chose a reference time slightly above the age of the star at engulfment to avoid to have zero for this difference which would imply an undetermined value when the logarithm is taken. 
The short-dashed line labeled $D_{orb}$ shows the evolution of the semi-major axis of the orbit of the planet in units of au. Superimposed to this short-dashed line, the analytical solutions to the orbit given by Eq. (\ref{equa:at}) is plotted as a red continuous line. 
%The curve labeled by a 1. is for a radius equal to 42.65 R$_{\odot}$, the lines labeled by a 2. and 3. are for radii equal to respectively 42.45 and 42.35 R$_{\odot}$. 
The dotted line labeled $D_{corot}$ is the corotation radius. The continuous lines (M$_{env}$) and long-dashed line (R$_{\star}$) show the evolution of the mass (in M$_{\odot}$) of the external convective envelope and of the radius of the star (in units of 40 R$_{\odot}$). The horizontal dotted lines (R$_{\star}$(ENG)) is the radius of the star at engulfment in units of au. 
{\it Upper right panel:} Acceleration of the reflex motion of the star as a function of the reflex motion of the star during the last 83300 y before planet engulfment. The vertical lines show the periods corresponding to different radii of the orbit. The dotted line on the right indicates the period just before engulfment (engulfment occurs when the radius of the orbit is equal to the stellar radius, which is 0.2 au).
The times in years correspond to the durations for passing from the orbit radius indicated to the left (see the labels of the vertical lines) to the radius indicated to the right of the times.
{\it Lower left panel:} Decrease of the stellar spin period as a function of the steallar spin period.
{\it Lower right panel:} Period variations during the planetary orbit-shrinking as a function of the orbital period. 
}
\label{figure:fall}
\end{figure*}

Red giant stars are a stellar evolutionary phase that is favorable
for studying the exchange of angular momentum
between a planetary orbit and its host star. The reason is that red giants are very slow rotators (most of them rotate more slowly
than 3 km s$^{-1}$ for surface gravities below about log $g\sim 2$), and
when they evolve along the red giant branch, they slow down. The observations of fast-rotating red giants
(typically with a rotation faster than 8 km s$^{-1}$, only a few percent are found) triggered the idea that these stars may have been spun up
by tidal interaction with a companion, possibly followed by the engulfment of that companion \citep{livio84, soker84, sackmann93, rasio96, siess99I, siess99II, villaver07, sato08, villaver09, carlberg09, nordhaus10, kunimoto11, bear11, mustill12, nordhaus13, villaver14, GIOI, GIOII}. 

In previous works, we studied the evolution of orbits of planets of different masses around stars of various initial masses and rotations with different orbital periods \citep{GIOI}. We determined the initial conditions required
for an engulfment to occur during the red giant phase. We also computed the evolution of the star after the engulfment \citep{GIOII} and explored the possibility that the fast rotation of the red giant after an engulfment might produce an observable surface magnetic field \citep{GIOIII}. In the present work, we focus on the short phase during which the orbit is shrinking before the planet engulfment. During that phase,  the red giant rotates
ever faster, the orbital period decreases, and the reflex motion of the star accelerates. We wish to obtain some orders of magnitudes for these changes and see whether they are of sufficiently high amplitude during reasonable periods of times to be observed,
eventually. If this were the case, this would offer extremely interesting clues about the physics of tides in such systems.

In Sect.~2 we briefly recall the main ingredients of our computations. The orbital evolution of a planet of 15 Jupiter mass ($M_J$) that orbits a 2 M$_\odot$ star is presented in Sect.~3, together 
with the variations as a function of time of the reflex motion ($\upsilon_{\rm reflex,*}$), of the stellar spin period ($P_{\rm rot,*}$), and 
of the planetary orbital period ($P_{\rm orb}$).  In Sect.~4 we place these results into a broader perspective and give our conclusions.

\section{Stellar and orbital evolution models}\label{sec:2}

%In the present investigation we need to model the evolution of the star, to make some assumption on the structure of the planet and
%to follow the evolution of the orbit of the planet in order to know when the engulfment will take place. 
%We discuss below the physics used to model these three different aspects of the problem.

We used computations performed by \citet{GIOI} that follow the evolution of the star and the evolution of the planetary orbit
in a consistent way.
By consistent, we mean here that the changes in orbit influence the stellar rotation through a boundary condition that allows angular momentum
to be removed from or added to the star (depending on the sign of the tidal forces), and the change in planetary orbit accounts for the changes in the properties of the star (the tidal forces
typically depend on the radius of the star and the size of the convective envelope, among other factors; see Eq.~1 below). \citet{GIOI} provide all the details about the physical ingredients used to compute these models.

The evolution of the semi-major axis $a$ of the planetary orbit, which we assume to be circular ($e=0$) and aligned with the equator of the star, accounts for the changes in the mass of the planet (by accretion of wind material or through evaporation), of the star
(through stellar winds), for the effects of the circumplanetary material (drag forces), and for the tides \citep[see][]{zahn66,alexander76,zahn77,zahn89,livio_soker84,villaver09,mustill12,villaver14}.
When the tidal term dominates, other forces such as the drag forces play a negligible role, and we concentrate here on the tidal force.
%The evolution of the semi-major axis $a$ of the planetary orbit, that we suppose circular ($e=0$) and aligned with the equator of the star, is given by \citep[see][]{zahn66,alexander76,zahn77,zahn89,livio_soker84,villaver09,mustill12,villaver14}
%\begin{equation}
%\left(\frac{\dot{a}}{a} \right)=
%\underbrace{-\frac{\dot{M}_{\star}+\dot{M}_{pl}}{M_{\star}+M_{pl}}}_{\rm Term\  1}
%\underbrace{-\frac{2}{M_{pl}v_{pl}}\left[F_{fri} + F_{gra}\right]}_{\rm Term\  2}
%\underbrace{-\left(\frac{\dot{a}}{a} \right)_{t}}_{\rm Term\  3}
%\ \ \ \ ,
%\label{equa:evoorb}
%\end{equation}
%where $\dot{M}_{\star}=-\dot{M}_{loss}$, $\dot{M}_{loss}$ being the mass loss rate (here given as a positive quantity), $M_{pl}$ and $\dot{M}_{pl}$ are the planetary mass and the rate of change in the planetary mass, $v_{pl}$ is the velocity of the planet, $F_{fri}$ and $F_{gra}$ are
%respectively the frictional and gravitational drag forces, $(\dot{a}/a)_{t}$ is the term that takes into account the effects due to the tidal forces. In this subsection, we discuss the equations used for estimating the different terms. Their importance for computing the orbital evolution is discussed in Sect. \ref{sec:4}.
The evolution of the semi-major axis $a$ that is due to the tides is given by \citet{zahn66,zahn77,zahn89},
\begin{equation}
(\dot{a}/a)_{\rm t}=\frac{f}{\tau}\frac{M_{\rm env}}{M_{\star}}q(1+q) \left(\frac{R_{\star}}{a}\right)^{8}\left(\frac{\Omega_{\star}}{\omega_{\rm pl}}-1\right),
\label{equa:tides}
\end{equation}
with $f=1$, when $\tau<P/2$, and equal to $(P/2\tau)^2$ otherwise, where 
$P$ is the orbital period, $\tau$ is the eddy turnover timescale, which is taken as in \cite{rasio96}.
This factor allows us to consider only the convective cells that contribute to the viscosity \citep{villaver09}.
$M_{\rm env}$ is the mass of the stellar convective envelope, $M_{\star}$ is the stellar mass, 
$q=M_{\rm pl}/M_{\star}$ with $M_{\rm pl}$ is the mass of the planet, 
$\Omega_{\star}$ is the angular velocity at the surface of the star, and 
$\omega_{\rm pl}$ is the orbital angular velocity of the planet.
%\begin{equation}
%\tau=\left[\frac{M_{env}(R_{\star}-R_{env})^{2}}{3L_{\star}} \right]^{1/3}\ \ \ \ ,
%\label{equa:tau}
%\end{equation}
%with $R_{env}$ the radius at the base of the convective envelope of the star.

The inverse of the above equation gives a timescale that corresponds to the time for 
$a$ to change by a factor $e$.  

\section{Changes in reflex movement, spin, and orbital periods before planet engulfment}

%\section{Amplitudes of the forces impacting the orbit of the planet}\label{sec:4} 
%We discuss in this section the orders of magnitudes of the various quantities involved in the computation of the evolution of the orbit of the planets (Eq.~\ref{equa:evoorb}). To do so 
We focus on a 15 M$_{\rm J}$ mass planet orbiting a 2 M$_{\odot}$ star at an initial distance of 1 au\footnote{By initial distance, we mean at a distance of 1 au on the ZAMS.}. We consider the case where the initial rotation of the star is equal to $\Omega/\Omega_{\rm crit}=0.1$ (i.e., an initial velocity on the ZAMS equal to 32.2 km s$^{-1}$). \citet{GIOI} showed that for such a system planet engulfment occurs during the red giant phase. 
%We provide in Table \ref{table:data_ZAMS_ENG} a few properties of the stellar model at the moment of engulfment. 

%\begin{table}
%\scriptsize{
%\caption{Duration for passing from a radius of the
%planet orbit R$_1$ to a radius R$_2$
%starting at a time about 63 Myr before the engulfment.}
% title of Table
%\resizebox{18cm}{!} {
%\begin{tabular}{ccc} 
%\hline\hline % inserts double horizontal lines
%R$_1$ & R$_2$ & $\Delta t$ \\
%$[$AU$]$&[AU]& [1000 yr] \\
%\hline % inserts single horizontal line
%1.0 & 0.9 & 58859 \\
%0.9 & 0.8 & 3414 \\
%0.8 & 0.7 & 760 \\
%0.7 & 0.6 & 234 \\
%0.6 & 0.5 &  63 \\
%0.5 & 0.4 & 17 \\
%0.4 & 0.3 & 3 \\
%0.3 & 0.2 & 0.3\\
%\hline\hline % inserts double horizontal lines
%\end{tabular}
%}
%}
%\label{table:falltime}
%\end{table}

%\subsection{Impact of tides}

%As we shall see, the term 3 is clearly the leading term in Eq. (\ref{equa:evoorb}), and we begin our discussion by a brief analysis of this term.
%In the present work, we consider the effects of tides only during the red giant phase, when a well developed external convective zone is present (see Eq.\ref{equa:tides}).
%Under the adopted prescription from Eq. (\ref{equa:tides}), we note that this term becomes effective only when a convective envelope is present, thus it is equal to zero during the whole pre-RG phase.

We show in Fig.~\ref{figure:fall} (upper left panel) the temporal evolution of a few quantities during the last 100 My before planet engulfment.
%for a phase comprising the 100 Myr before engulfment. 
In particular, the dashed line shows the evolution of the semi-major axis of the orbit of the planet (${\rm D}_{\rm orb}=a$). 
%This curve was computed following the evolution of the stellar characteristics as the evolution of the radius, of the mass of the exterior convective envelope and so on...
%%We see that the appearance of the external convective envelope ({\it i.e.} when $M_{\rm env}$ becomes non-zero) is concomitant with an inflation of the star. This is because inflation increases the opacity in the external layers and thus favors convection. This inflation lowers considerably the surface angular velocity of the star pushing out the corotation radius (D$_{\rm corot}$), which becomes, from this time on, much larger than the actual orbital radius (D$_{\rm orb}$)\footnote{On the ZAMS, the orbital period of the planet (0.7 year) is much larger than the rotation period of the star (7.5 hours), thus the planet orbits well outside the corotation radius. The corotation radius on the ZAMS is 0.000038 au, much smaller than the semi-major axis of the planet orbit. The corotation radius is even smaller than the radius of the star which is 1.19 R$_{\odot}$ on the ZAMS, i.e. 0.00555 au.}.
%%This is why, from this time on, the tidal forces will make the orbital radius to decrease as a function of time (inward migration). The shrinking of the planetary orbit accelerates as a function of time. 
%This is shown in Table \ref{table:falltime}. 
To better understand what occurs, we adopt a simplified equation for the tidal torque given in Eq.~\ref{equa:tides}:

\begin{equation}
a^7(\dot{a})_{t}=- \underbrace{
\frac{f}{\tau}\frac{M_{env}}{M_{\star}}q(1+q) R_{\star}^{8}
}_{C}\ \ \ \ .\end{equation}
To obtain this equation, we assume that $\left(\frac{\Omega_{\star}}{\omega_{pl}}-1\right) \sim -1$, that is, {\it } $\Omega_{\star}$ is significantly smaller than $\omega_{pl}$, or $P_{\rm rot}$ is longer than $P_{\rm orb}$.
This is indeed the case. Compare the abscissas of the two lower panels in Fig.~1. $P_{\rm rot}$ is a factor 3-5 longer than $P_{\rm orb}$ during the phase shown. 
%Actually forgetting this approximation is numerically easy, and it did not produce very important changes. So we keep this simplification here. 
%It
%would change a little bit the durations of the phases during which a given variation can be observed by amount that would be between 20-33\%. This is of course not negligible
%but for not tremendoulsy important for the purpose of this exploratory works.
%We checked that accounting for the $\left(\frac{\Omega_{\star}}{\omega_{pl}}-1\right)$ 
%shifts the time of the engulfment by about one thousand year, but does not change otherwise the overall shape of the
%curve. In this work we are not so much interested to make a prediction for the precise time when an engulfment will occur, but to estimate the amplitudes of the time variations
%of quantities related to the change of the orbit and for how long these variations are above some limits.
During the last 100 000 years before engulfment, the right-hand term can be considered constant. In that case, the equation has an analytical solution given by

\begin{equation}
a(t)=(a_{\rm 0}^8-8C(t-t_0))^{1/8},
\label{equa:at}
\end{equation}
where $a_{\rm 0}$ is the radius of the orbit at the time $t_0$. 
%From Fig.~1, we see that during the last 100 000 years, the characteristics of the star keep nearly constant values. 
By adopting values obtained at the time of engulfment as representative values for this whole period and plugging them into the analytic solution, 
%Let us adopt the properties of the star
%at engulfment (see Table \ref{table:data_ZAMS_ENG}). 
we obtain the (red) continuous line in Fig. \ref{figure:fall}. There is a slight mismatch with the solution obtained in the complete numerical model following the evolution of the star and of the orbit
(see the black dashed line) that is due mainly to the fact that stellar evolution computations prevent us from reducing the time
steps as much as we would like. The analytic solution, on the other hand, while 
providing an excellent fit to the more complete numerical solution, allows us to describe the very last phases before the engulfment in much more detail.
The shrinking of the planetary orbit strongly accelerates due to the ever increasing tidal forces when the
distance to the star decreases.

In the upper right panel of Fig.~\ref{figure:fall},  the change in the reflex motion of the star is plotted as a function of the reflex velocity. The reflex velocity is given by $\upsilon_{\rm reflex,*}=\sqrt{G M_*/a}\  m_{\rm pl}/M_*$, with $G$ the gravitational constant. It can be written $\upsilon_{\rm reflex,*} \approx 30 \sqrt{M_*/a}\  m_{\rm pl}/M_*$ km s$^{-1}$ , where the masses are in solar units and $a$ in au. For our case,
this formula typically gives a velocity of around 0.580 km s$^{-1}$ when $a$ is equal to 0.3 au. The acceleration of this reflex motion is given by ${\rm d}\upsilon_{\rm reflex,*}/{\rm d}t=-0.5\ \upsilon_{\rm reflex,*}\ \dot{a}/a$. 
The inverse of $\dot{a}/a$ is a time. An approximate value of it can be read in the right upper panel of Fig.~1,
for instance, from the durations that are indicated at the top of the panel. The timescale is typically around 1500 y for $a$ equal to 0.3 au, which translates into an acceleration of the order of 20 cm s$^{-1}$ per year. 
During the four decades before the engulfment, this acceleration reaches values higher than one meter per second per year. However, as discussed in Sect.~4, such low values are beyond reach because
they are blurred by
the jitter that is caused by convection in red giants.
%While these quantities are small, they are not out of reach of present facilities that
%can reach radial velocity measurement with a precision of the order of the meter per second \citep{Mayor2003} and of future ones as ESPRESSO on the VLT that will reach precision of the order of 10 cm s$^{-1}$ \citep{Pepe2013}.
%Let us note however red giant have a large convective outer zone with velocities of the convective cells that can recah values up to a few km s$^{-1}$, making this kind of observations very difficult if not impossible.

%Let us recall here that only a few percents of red giants show so fast surface rotation velocities that an interaction with a companion has to be invoked. The duration of the red giant branch phase is of the order of 100 My for a 2 M$_\odot$. This means that if only 2\% of the 2 M$_\odot$ will go through an engulfment of a massive planet, the chance to pick one in that stage among 2 M$_\odot$ stars would be 0.02 40/10$^8$ $\sim$ 8 10$^{-9}$, thus extremely low. 
%decrease of the spin period

The variation in stellar spin is shown in the lower left panel of Fig.~1. The rotation period varies by more than about 10 minutes per year in the 3000 years preceding the engulfment.
This corresponds to a relative change in the rotation period (around 0.65 y during that period) of more than 3 10$^{-5}$. 
This low value is beyond reach of current observations (see Sect.~4). 

%This is again a quite small value, likely out of reach without very long baseline of observations. Moreover,
%3000 y remains a short duration compared to the red giant lifetime. Using the same numbers as in the previous paragraph, the chance to pick one star in that stage among 2 M$_\odot$ stars would be 0.02 3000/10$^8$ $\sim$ 6 10$^{-7}$!
%So also extremely low.

%decrease of the orbital period
The most promising way of detecting the effects of the tides would probably be through the change in orbital period.
The variation in orbital period is shown
in the lower right panel of Fig.~\ref{figure:fall} for the last part of the evolution of the orbit. We see, for instance, that when the orbital period is around 0.18 year, about 3000 years before the engulfment, the period change is about 5 minutes per year.
Observing such a star for five years would accumulate data for nearly 28 revolutions, and the change between the first and the last orbital period would be of about 25 minutes.
Such a period difference is detectable for transiting planets for which accurate measurements of the variation of the beginning
or end of the eclipse can be obtained. 

How do these estimates change when the mass of the star, its rotation rate, the mass of the planet, and its initial distance to the star vary?
It is beyond the scope of this letter to discuss these dependencies in details. This will be the topic of a more extended work. We
content ourselves here with a few general remarks.
%\begin{itemize}
%\item 
It can be shown that the derivatives of $\upsilon_{\rm reflex,*}$, $P_{\rm spin}$ and $P_{\rm orb}$ are proportional to $\dot{a}/a,$ which is the inverse for the timescale of the orbit shrinkage.
Thus the higher $\dot{a}/a$, the larger the derivatives, but on a shorter period. This is independent of the system characteristics.
%\item 
The quantity $\dot{a}/a$  is dominated by the term in $R_*^8$, thus by the radius of the star at the time of the engulfment. The time of the engulfment and thus this radius depends
on the initial conditions considered. For a given initial mass of the star, the engulfment occurs later and thus at larger radius along the red giant branch for less massive planets with larger initial orbits \citep[see Table A.1 in][]{GIOI}. Thus the largest variations are expected for these systems. We note that the time of the engulfment depends weakly on the initial rotation of the star.
%\item 
$\dot{a}/a$ varies with $q,$ the ratio between the mass of the planet to that of the star (since during the red giant phase
$M_{\rm env}$ is about $M_*$, we consider here that the factor $M_{\rm env}/M_*$ is about 1). Thus everything being equal, the variations are largest when $q$ is large. However, as indicated above, when $q$ is large,
the engulfment occurs at earlier time when the red giant radius is smaller, making the variations smaller. This last effect in general dominates, and we therefore expect to see the largest variations
for lower $q$ values.
%\end{itemize}
This means that the largest variations are expected for systems whose orbit shrinkage occurs near the top of the red giant branch. This shrinkage occurs near the top for systems with low $q$ values and
for which the initial distance of the planet to the star is large. 

\section{Discussion and conclusion}

Two great challenges need to be met to have a chance to see the variations discussed in this paper. First, we need very precise measurements at different times of the spin or orbital periods or of the reflex motion. Second, we need a careful selection of the red giants that are worthwhile to be  monitored.

We discuss the first point.
We have seen above that during the four decades before the engulfment, the reflex motion accelerates at a rate of more than one meter per second per year. 
While these quantities are small, they are not beyond reach of present facilities that
can reach radial velocity measurement with a precision of the order of one meter per second \citep{Mayor2003} and of future facilities such as ESPRESSO on the VLT, which will reach a precision of the order of 10 cm s$^{-1}$ \citep{Pepe2013}.
We note, however, that red giants have a large external convective zone with velocities of the convective cells that can reach values
of up to a few km s$^{-1}$, which means that the detection of such an effect is probably beyond reach\footnote{It should be noted, however, that convection produces a constant random noise, while the effect studied here consists of a shift.}. 

The change in spin period of the star is of the order of 10 min per year in the last 3000 y before engulfment. The rotation period is shorter than 0.65 y during this period, which means that observing the star for five years would collect data for about 10 rotations, and a change of nearly one hour would be found between the first and last spin period. This corresponds to a relative change ($\Delta P_{\rm spin}/P_{\rm spin}=-\Delta  \upsilon_{\rm eq}/\upsilon_{\rm eq}$, where $\upsilon_{\rm eq}$ is the equatorial rotation velocity) of 2.3 10$^{-5}$.
This is again a very delicate observation to be performed. Note that the velocity of the convective cells is of the same order of magnitude as the equatorial velocity and contributes to a similar amount to the enlargement of the absorption lines.

As we indicated above, the change in orbital period of a transiting planet is the most promising way of detecting the effect discussed here. Using transiting planets means that very accurate determinations of the period changes can be obtained, observing shifts of the beginning and/or end of the transit. The typically expected time shifts are of the order of 
about one hour or more
in a time span of about five years during the last $\sim$3000 y before the engulfment.
%The reasons for privileging changes of the orbital period are that the orbital periods are shorter than the spin periods by a factor 3 to 5. Second, using transiting planets, it means that very accurate determinations of the period changes can be obtained. 
Some first period variations have been observed for planet-orbiting subgiants
\citep{Macie2016, Patra2017}\footnote{We note that for these two cases, the authors
tend to attribute the observed period changes to the tidal effects, although they do not exclude other causes.}. The case of subgiants is more favorable for detecting such effects. Subgiants have smaller radii, thus the planets can orbit with shorter orbital periods than around red giants. This  allows accumulating many
more orbital revolutions in a given time frame. 
On the other hand, at the subgiant phase, the signs of the star being a fast rotator as a result of some tidal interactions are more difficult to detect since during this phase the surface rotation of the star still depends strongly on the initial
rotation \citep[see Fig.~1 in][] {GIOI}. This complicates the identifications of the interesting candidates. More evolved red giants, regardless of the initial rotation of the star, rotate so slowly that any rotation above some level
%let us say 3-4 km s$^{-1}$ 
might be used to identify candidates that are interesting for follow-up observations.

%a similar observation around red giants, although more challenging, would be extremely interesting. First, it would allow
%to test the tidal force when a large outer convective zone is present. Second, since regular red giants rotate so slowly, a rotation above some level might be used to identify interesting candidates for follow-up observations (see below). 

We now consider the selection of candidates to be monitored.
Candidates that would fall in the frame of the computations shown here would present the following features: the host star should be a red giant with a log $g$ below about 2 and a surface rotational velocity higher than 3-4 km s$^{-1}$. A planet with at least one Jupiter mass should have been detected to orbit the star at a distance
smaller than 1 au. There is some chance that such a system would be in the orbit-shrinkage phase, unless the high observed velocity were due to a previous planet engulfment. When we discard this last possibility and assume a duration for the orbit shrinkage of about 100 000 years, the period during which the variation
amplitudes are about one hour or more over an observing time of five years
would be about 3000 years, which means about 3\%.  This means that of 1000 red giants with the properties above, very roughly 30 should be in the shrinking period. The transit probability can easily be derived and is equal to  $\sim R_{\star}/a$. In our case it amounts to 40/214$\sim$ 18.7\%, which means that  between five and six might show a transit. Of course, the larger the star, the more luminous it is, and thus also the more difficult it is to detect a transit.

About 100 red giants are known (the precise number is 99) that show the presence of a planet\footnote{see the catalog on the web https://www.lsw.uni-heidelberg.de/users/sreffert/giantplanets/giantplanets.php}. 
Nine of these consists of a star with a mass between 1.8 and 2.2 M$_\odot$, that is, in the range of the case studied here, and have a planet at a distance shorter than 1 au. There is no case at the moment
that would show both features (log $g$ smaller to 2 with $\upsilon_{\rm eq}$ higher than 3-4 km s$^{-1}$), which would be indicative of planetary orbit shrinking. 
A very interesting case is TYC 3667-1289b \citep{Nied2016} ($\sim$1.9 M$_\odot$ star with a planet of 5.4 M$_J$ at least orbiting at a distance of 0.21 au). 
Interestingly, it has a chance of about 14\% to present a transit. On the other hand, the rotation velocity of the star is not very high with a $\upsilon\sin i$ of 3.2 km s$^{-1}$\footnote{If transit is detected and we have a coplanar planet, then this mean that this value is near the true equatorial velocity.}, therefore there is little chance that the planet is in a phase of fast orbital shrinking. However, it might be an interesting object to continue to observe, especially if it shows a transit.
%with a planet at a distance inferior to 1 au (31), we did not find any case with a star mass 
%showing the two features (log $g$ inferior to 2 with $\upsilon_{\rm eq}$ larger than 3-4 km s$^{-1}$) that would be indicative that the planetary orbit might be shrinking. 
%%Of course it does not mean that such systems does not exist but more likely 
%%that we have to wait for more data to have a chance to detect such a favorable case. Also, the present analysis should be extended to cases of planet engulfment occurring at earlier stages of the evolution of the star.}
%Probably a useful diagram in that respect would be to place red giant host stars with a detected planet in a diagram
%where $D_{\rm orb}$ is plotted as a function of the surface gravity of the star. In such a diagram, planets in non-shrinking orbit should be all above a line showing the minimum distance that the planet should be
%from the star to avoid any engulfment during the red giant phase. Any system detected below such a limit (which is somewhat model dependent however) would represent an interesting candidate. The condition is however not 
%sufficient. In addition, the red giant should present some rotation rate above 3-4 km $^{-1}$ so that it will show some sign of having been accelerated by the tidal force. 
%It is difficult at the present time to say whether such systems would be discovered one day, but at least it does not appear very unrealistic that with more data, once, a system would be observed in its
%orbit shrinking phase during the red giant phase. 
As we mentioned, the study of such systems would bring very interesting and probably unique constraints on the physics of tides. Measuring the timescale of the orbit shrinking for a few systems would allow us in particular to check the dependance of this timescale on the radius of the star, an important first step for assessing the physics of tides. This might be an interesting science case for a space mission like PLATO, which will be dedicated to transiting planets.

%what is the accuracy needed?

%How does it depend on the specific case considered here?

%*How to  detect candidates?

\begin{acknowledgements} 
%This research has made use of the Exoplanet Orbit Database
%and the Exoplanet Data Explorer at exoplanets.org. 
We thank the referee Eva Villaver for very valuable suggestions.
The project has been supported by Swiss National Science Foundation grant 200020-172505.
This work has been carried out in part within the frame of the National Centre for Competence in Research PlanetS
supported by the Swiss National Science Foundation. 
%AAV acknowledges support from an Ambizione Fellowship of the Swiss National Science Foundation.
\end{acknowledgements}

\bibliographystyle{aa} % style aa.bst
\bibliography{biblio} % your references Yourfile.bib

\end{document}